\documentclass[iop]{emulateapj}
\usepackage{graphicx}
\usepackage{setspace}
\usepackage{natbib}
\usepackage{color}
\usepackage{amsmath,amssymb}
\usepackage{times}
\usepackage{aas_macros}

\begin{document}

\title{Elliptical galaxy masses out to five effective radii: the realm of dark matter}
\author{A.J Deason, V. Belokurov, N.W. Evans and I.G. McCarthy\altaffilmark{1}}
\affil{Institute of Astronomy, Madingley Rd, Cambridge, CB3 0HA}

\altaffiltext{1}{Kavli Institute for Cosmology, University of Cambridge,
  Madingley Road, Cambridge, CB3 OHA}

\date{\today}

\begin{abstract}
  We estimate the masses of elliptical galaxies out to five effective
  radii using planetary nebulae and globular clusters as tracers. A
  sample of 15 elliptical galaxies with a broad variation in mass is
  compiled from the literature. A distribution function-maximum
  likelihood analysis is used to estimate the overall potential slope,
  normalisation and velocity anisotropy of the tracers. We
    assume power-law profiles for the potential and tracer density and
    a constant velocity anisotropy. The derived potential power-law
  indices lie in between the isothermal and Keplerian regime and vary
  with mass: there is tentative evidence that the less massive
    galaxies have steeper potential profiles than the more massive
    galaxies. We use stellar mass-to-light ratios appropriate for
  either a Chabrier/KTG (Kroupa, Tout \& Gilmore) or Salpeter initial
  mass function to disentangle the stellar and dark matter
  components. The fraction of dark matter within five effective radii
  increases with mass, in agreement with several other studies. We
  employ simple models to show that a combination of star formation
  efficiency and baryon extent are able to account for this
  trend. These models are in good agreement with both our measurements
  out to five effective radii and recent SLACS measurements within one
  effective radii when a universal Chabrier/KTG initial mass function
  is adopted.

\end{abstract}

\section{Introduction}

There is strong evidence for the presence of dark matter in spiral
galaxies where the rotation curves of their extended cold gas discs
remain flat out to large radii. However, elliptical galaxies are
generally free of cold gas so they lack an equivalent to the HI
rotation curves of spiral galaxies. Furthermore, the stellar component
of ellipticals is dominated by random motions so their kinematics are
more difficult to model and the results are bedevilled by the
mass-anisotropy degeneracy.

The confirmation of dark matter haloes surrounding elliptical galaxies
has largely been confined to the brightest galaxies using either the
X-ray emission of their hot gas (e.g. \citealt{loewenstein99};
\citealt{osullivan04}; \citealt{humphrey06}; \citealt{johnson09};
\citealt{das10}) or strong lensing techniques (e.g. \citealt{treu04};
\citealt{rusin05}; \citealt{gavazzi07}; \citealt{koopmans09};
\citealt{auger10b},\citealt{faure11},\citealt{leier11}). Stellar
dynamical studies from integrated light spectra can also be used to
estimate dynamical masses (e.g. \citealt{gerhard01};
\citealt{cappellari06}; \citealt{thomas07}; \citealt{tortora09}), but
such studies are generally limited to within a couple of effective
radii, $R_{\rm eff}$.

To study the outer reaches of elliptical galaxies (i.e. beyond $\sim
2$ effective radii) requires distant tracers such as planetary nebulae
(PNe) or globular clusters (GCs). In particular, PNe can be used to
trace intermediate mass galaxies, whereas GCs and other mass probes
are generally biased towards more massive systems. To make use of the
larger mass range probed by the PNe, \citet{douglas02} developed a
specialised instrument -- the PNe Spectrograph (PN.S) -- to study the
kinematics of these tracers in elliptical galaxies.  The early results
of this project suggested a dearth of dark matter in ordinary
ellipticals (\citealt{romanowsky03}). However, \cite{dekel05} showed
that a declining velocity dispersion profile is consistent with a
massive dark halo if the tracer anisotropy is radially biased. More
recent work utilising PNe to trace the kinematics of intermediate mass
ellipticals find that although there is some evidence for the presence
of dark matter haloes, the fraction of dark matter (within $5R_{\rm
  eff}$) is somewhat lower than their higher mass analogues
(e.g. \citealt{douglas07}; \citealt{napolitano09};
\citealt{napolitano11}). In addition, \cite{napolitano09} find that
the dark matter halo concentrations (i.e $c_{200}$ or $c_{\rm vir}$)
of these intermediate mass ellipticals are lower than $\Lambda$CDM
predictions. However, estimating the dark matter halo mass and
concentration at the virial radius requires a large extrapolation from
the radial range of the current data ($5R_{\rm eff} \sim 0.1r_{\rm
  vir}$). In fact, \cite{mamon05} caution that extrapolation of
dynamical studies within $\sim 5 R_{\rm eff}$ to the virial radius are
fraught with large uncertainties. A comparison between intermediate
and high mass ellipticals is difficult as it is rare for both mass
scales to be probed using the same method.

Whilst the overall mass of a galaxy can be derived using dynamical
modelling of tracer kinematics, one must disentangle the stellar
component in order to study the properties of the dark matter
halo. However, incomplete knowledge of the initial mass function (IMF)
inhibits this decomposition. While studies of our own Galaxy favour an
IMF with a flattened slope below $0.5M_\odot$ (\citealt{scalo86};
\citealt{kroupa93}; \citealt{chabrier03}), near infrared spectroscopic
studies of massive ellipticals find that the low mass slope of the IMF
may become steeper (\citealt{vandokkum10}). Furthermore, several
recent studies have suggested a mass-dependent IMF
(e.g. \citealt{auger10a}; \citealt{treu10}).

\begin{table*}
\centering
\renewcommand{\tabcolsep}{0.04cm}
\renewcommand{\arraystretch}{1.5}
\begin{tabular}{| c c c c c c c c c c c c c c r |}
\hline
Name  & Type & Tracer & D & Ra & Dec & PA & $R_{\rm eff}^\dagger$ & $L_{B}$ & $V_h$ &
$\alpha^\star$ & KS & $N(>2R_{\rm eff})$ & $\left(V_{\rm rot}/\sigma\right)^2$ & References\\
(NGC) & & & [kpc] & [deg] & [deg] & [deg] & [asec] & $10^{10}
L_{B,\odot}$ & kms$^{-1}$ & & & & &\\
\hline
0821 &  E6 & PNe & 22.4 & 32.1 & 11.0 & 25 & 39 & 2.6 & 1735 &
3.3 &  1.0 & 55 & 0.10 & Coc09$^{\dagger \star}$\\
1344 & E5 & PNe & 18.9 & 52.2 & -31.2 & 167 & 46 & 1.8 & 1169 &
3.3 & 1.0 & 86 & 0.04 &T05$^\dagger$;Coc09$^\star$\\
1399 & E1 &  GC & 19.0 & 54.6 & -35.5 & 110 & 42 & 3.5 & 1442 &
2.7 & 0.6 &  444 & 0.01 & Sag00$^\dagger$;Sch10$^\star$\\
1407 & E0 &  GC & 20.9 & 55.1 & -18.6 & 60 & 57 & 4.7 & 1784 &
2.6 & - & 134 & 0.04 & F06$^\star$; R09$^\dagger$\\
3377 &  E5 & PNe & 10.4 & 161.9 & 14.0 & 35 & 41 & 0.6 & 665 &
3.2 & 1.0 & 81 & 0.14 & Coc09$^{\dagger \star}$\\
3379 &  E1 & PNe & 9.8 & 162.0 & 12.6 & 70 & 47 & 1.4 & 889 & 3.3
& 1.0 & 89  & 0.03 & Dou07;Coc09$^{\dagger \star}$\\
4374 &  E1 & PNe & 17.1 & 186.3 & 12.9 & 135 & 72.5 & 5.0 & 1060 &
3.1 & 1.0 & 234 & 0.03 & Coc09$^\star$; N11$^\dagger$\\
4486 &  E0 & GC & 17.2 & 187.7 & 12.4 & 160 & 105 & 7.3 & 1350 &
2.4 & 0.3 & 157 & 0.18 & Cot01$^\star$; Cap06$^\dagger$; Mei07 \\
4494 &  E1 & PNe & 15.8 & 187.9 & 25.8 & 0 & 53 & 2.2 & 1344 &
3.4 & 0.9 & 108 & 0.03 &Coc09$^{\dagger \star}$; N09\\
4564 &  E6 & PNe & 13.9 & 189.1 & 11.4 & 47 & 22 & 0.5 & 1142 &
3.4 & 0.3 & 29 & 1.00 &Coc09$^{\dagger \star}$\\
4636 &  E0 & GC & 15.0 & 190.7 & 2.7 & 145 & 108 & 2.7 & 906 &
2.6 & - & 105& 0.02 & Dir05$^\star$; Sch06$^\dagger$\\
4649 &  E2 & GC/PNe & 17.3 & 190.9 & 11.6 & 105 & 110 & 6.1 &
1117 & 2.8/3.2 & 1.0/0.8 & 61/68 & 0.42/0.04 & H08$^\star$; L08$^\dagger$; T11$^\star$\\
4697 &  E6 & PNe & 10.9 & 192.2 & -5.8 & 70 & 66 & 1.9 & 1241 &
3.4 & 1.0 & 180 & 0.10 & De08$^\dagger$;Coc09$^\star$;Men09\\
5128 &  E/S0 & GC/PNe & 3.8 & 201.4 & -43.0 & 35 & 300 & 2.7 & 541
& 3.4/3.5 & - & 156/323 & 0.10/0.38 & W10$^{\dagger \star}$; P04\\
5846 &  E0 & PNe & 23.1 & 226.6 & 1.6 & 70 & 81 & 4.0 & 1714 &
3.1 & 1.0 & 55 & 0.01 &Cap06$^\dagger$;Coc09$^\star$\\
\hline
\end{tabular}
\caption[]{Parent galaxy properties - (1) Name of galaxy. (2) Type of
  tracer (GC or PNe). (3) Distance in Mpc. (4) and (5) Right ascension
  and declination. (6) Position angle. (7) Effective radius. (8) B
  band luminosity. The B band luminosity is derived from the
  \cite{devac01} apparent magnitudes and corrected for extinction
  according to \cite{schlegel98}. (9) Heliocentric velocity. (10)
  Single power-law density index of tracers.(11) KS probability of
  approximate single power-law begin drawn from observed density model (12) Number of tracers
  beyond two effective radii. (13) Fraction of kinetic energy in
  rotation.(14) References. The source of the effective radii and
  tracer density distribution are indicated by the $\dagger$ and
  $\star$ symbols respectively: \citealt{cappellari06} (Cap06),
  \citealt{coccato09} (Coc09), \citealt{cote01} (Cot01),
  \citealt{delorenzi08} (De08), \citealt{dirsch05} (Dir05),
  \citealt{douglas07} (Dou07), \citealt{forbes06} (F06),
  \citealt{hwang08} (H08), \citealt{lee08} (L08), \citealt{mei07}
  (Mei07), \citealt{mendez09} (Men09), \citealt{napolitano09} (N09),
  \citealt{napolitano11} (N11), \citealt{peng04} (P04),
  \citealt{romanowsky09} (R09), \citealt{saglia00} (Sag00),
  \citealt{schuberth06} (Sch06), \citealt{schuberth10} (Sch10),
  \citealt{teodorescu05} (T05), \citealt{teodorescu11} (T11),
  \citealt{woodley10} (W10)}
\label{tab:parent}
\end{table*}

Clearly, our knowledge of the dark matter haloes surrounding
elliptical galaxies is far from complete. In this work, we study the
outer regions of elliptical galaxies over a range of masses.  These
regions are relatively unexplored, especially for the less massive
systems. To this end, we compile a sample of galaxies with kinematic
tracers beyond $\sim 2R_{\rm eff}$. Previous authors have used either
Jeans modelling (e.g. \citealt{napolitano09}; \citealt{napolitano11})
or orbit library techniques, such as Schwarzschild
(e.g. \citealt{thomas11}) or NMAGIC modelling (e.g
\citealt{delorenzi09}), to study the kinematics of such
tracers. Whilst effective, the latter methods are arduous and
generally applied on a galaxy by galaxy basis. In the coming years,
where the number of kinematic tracers surrounding elliptical galaxies
is likely to dramatically increase, it is important to develop methods
to analyse a large sample of systems both quickly and
effectively. Herein, we adopt a distribution function analysis. The
advantage of such a scheme is that it allows us to study a number of
systems with relative ease so the overall trends with mass can start
to be addressed.

The paper is arranged as follows. In \S2, we introduce our sample of
elliptical galaxies with distant kinematic tracers compiled from the
literature. \S3 describes the distribution functions and maximum
likelihood analysis. We give our results in \S4 and develop some
simple model predictions. Finally, we draw our main
conclusions in \S5.

\section{Elliptical galaxy sample}

Our aim is to probe the dark matter haloes of early type galaxies. To
this end, we construct a sample of local galaxies with tracers
reaching beyond two effective radii. Our sample of 15 galaxies is
compiled from the literature -- the properties of these systems are
given in Table \ref{tab:parent} along with the associated
references. This sample covers a range of galaxy masses and
environments (i.e. from field to cluster galaxies). The tracers are
either GCs or PNe. In two cases (NGC 4649 and NGC 5128),
we use both GCs and PNe as tracers. As the different
tracers may have different dynamical properties (e.g. anisotropy), we
analyse each sample separately. However, we do not attempt to model
red and blue globular cluster populations separately. Whilst previous
authors have found these populations may have different density
profiles and orbital properties (e.g. \citealt{cote01};
\citealt{hwang08}; \citealt{schuberth10}), we choose to study the
globular cluster population as a whole to maximise the number of
tracers. Many of the PNe samples derive from the PNe Spectrograph
project\footnote{http://www.strw.leidenuniv.nl/pns/PNS\_public\_web/PN.S\_project.html}. This
project specifically targets the outer regions of local galaxies and
promises to increase substantially the number of systems with
dynamical tracers in the near future.

We exclude any obvious outliers in the samples using a 3-$\sigma$
velocity clipping method (see e.g. \citealt{douglas07}). The line of
sight velocities of tracers at similar projected radii are used to
exclude any objects with obviously inflated velocities (i.e. those
with velocities exceeding 3$\sigma$). In addition, we exclude outliers
flagged in the literature. For example, we exclude those PNe from the
NGC 3379 sample identified by \cite{douglas07} as belonging to NGC
3384.

For each sample of tracers, we approximate the density profile by a
single power-law $\rho \propto r^{-\alpha}$. We use the density
profiles given in the literature. Where 2D profiles are given, we fit
a single power-law to the projected distribution and increase the
power-law index by 1 to convert to a 3D spatial profile. The
approximate single power-law is only fitted between the range of radii
we are probing (e.g. between $2R_{\rm eff}$ and $R_{\rm max}$). In
Table \ref{tab:parent}, we give the Kolmogorov-Smirnov (KS)
probability that our approximate single power-law is drawn from the
density profiles given in the literature. This statistic is omitted if
the literature profiles are given as single power-laws. In most cases,
a single power-law is a good approximation (with probability $\sim
1$). However, in NGC 4564 and NGC 4486 the probability is only $\sim
0.3$. Inspection of these profiles shows that the disagreement is
driven by the outer 10\% of tracers and hence the approximate single
power-law is a good representation of 90\% of the sample\footnote{Note
  that an increase (or decrease) of the tracer density power-law index
  by 1 dex leads to an increase (or decrease) in the mass estimate of
  $\sim 30\%$. On average, our mass estimates are known to 20\% (see
  Section 4); hence, only if the power-law is changed by 1 dex (a
  gross overestimate of the uncertainty) can the masses increase or
  decrease by more than the statistical errors.}. In this exercise, we
use the derived density profiles in the literature rather than the
published positional data. Many of these density models are derived by
taking into account completeness and selection biases
(e.g. \citealt{dirsch05}; \citealt{douglas07}; \citealt{schuberth10})
and this information is not as readily accessible as the published
positional and velocity data. However, we have also performed KS tests
relative to published tracer positions and find similar results.

Most tracers have a density power-law in the range $\alpha =2-4$. In
general, the GCs have more extended distributions than the PNe. This
may be related to the parent galaxy properties as the GCs tend to
trace the more massive galaxies. However, this may also be due to
differences between the GC and PNe populations themselves. In general,
PNe tend to follow the light distribution (e.g. \citealt{coccato09},
but see \citealt{douglas07}), but the GCs can often be more extended
(e.g. \citealt{dirsch05})

\begin{figure}
  \centering
    \includegraphics[width=3in,height=3.75in]{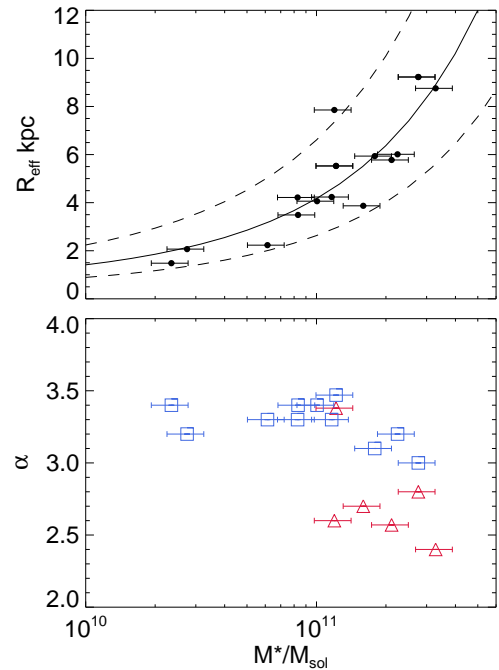}
    \caption[0]{Top panel: Size-mass relationship for the elliptical
      galaxy sample. The solid line shows the quadratic model of \cite{hyde09}
      and the dashed lines indicate a scatter of 0.2dex per stellar
      mass bin. Here, a Chabrier/KTG IMF is adopted. Bottom panel:
      Tracer power-law density index vs. stellar mass. The red
      triangles and
      blue squares indicate GCs and PNe respectively. Higher mass
      galaxies tend to have tracers with shallower density profiles.}
    \label{fig:size_mass}
 \end{figure}

In Fig. \ref{fig:size_mass}, we show the
size-mass relation of our sample (assuming a Chabrier/KTG IMF) and the
relation between tracer power-law index and stellar mass. As can be
seen in these panels, more massive galaxies tend to have more extended
tracer populations and larger effective radii. This size-mass relation
proves important to our conclusions and is discussed further in
Section \ref{sec:dm}. Note that the effective radii we have adopted
(given in Table \ref{tab:parent}) are generally consistent with de
Vaucouleurs $R^{1/4}$ fits. This eases comparison with studies like
\cite{hyde09} who adopt de Vaucouleur surface brightness
profiles. However, we caution that the effective radii can change
substantially if a general Sersic $R^{1/n}$ profile is adopted.

\section{Distribution Function}

We analyse the dynamical properties of the tracers using a
distribution function (henceforth DF) method. DFs are a valuable tool
for studying steady state-systems as they replace the impracticality
of following individual orbits with a phase-space probability density
function. We provide a brief description of these DFs below but direct
the interested reader to \cite{evans97} and \cite{deason11}, where
more detailed descriptions are given.

For simplicity, we use power-law profiles for the tracer density and
potential, namely $\rho \propto r^{-\alpha}$ and $\Phi \propto
r^{-\gamma}$, where $\alpha$ and $\gamma$ are constants.  Although our
formulae hold good for $\gamma \in [-1,1]$, models with $\gamma <0$
are less useful for modelling galaxies. The velocity distribution is
given in terms of the binding energy $E =
\Phi(r)-\frac{1}{2}(v_l^2+v_b^2+v_{\rm los}^2)$ and the total angular
momentum $L = \sqrt{L_x^2+L_y^2+L_z^2}$,
\begin{equation}
F(E,L) \propto L^{-2\beta} f(E) \rho(r)
\label{eq:even}
\end{equation}
where
\begin{equation}
\label{eq:df}
f(E) = E^{\beta(\gamma-2)/\gamma+\alpha/\gamma-3/2}
\end{equation}
Here, $\beta$ is the Binney anisotropy parameter (\citealt{binney80}), namely
\begin{equation}
\beta=1-\frac{\langle v^{2}_{\theta} \rangle+ \langle v^{2}_{\phi}
  \rangle}{2 \langle v^2_r \rangle},
\end{equation}
which is constant for the DFs of form eqn~(\ref{eq:even}).

Our analysis assumes spherical symmetry. While the majority of our
elliptical galaxies look spherical in projection (E0/E1) there are a
non-negligible number which are more flattened (E6). However, we only
consider tracers beyond $2R_{\rm eff}$, well beyond the region from
which the galaxy type was inferred. With no prior knowledge of the
shape of the potential in these regions, we make the simplest
assumption of spherical symmetry. Relaxing this assumption calls for
more sophisticated modelling (e.g. \citealt{delorenzi07}). However, we
note that it is not obvious whether this extra complication makes any
appreciable difference to the mass estimates (e.g
\citealt{delorenzi09}). 

These distribution functions can easily be adapted to probe the
rotational properties of the tracers (see
\citealt{deason11}). However, in the spherical approximation, it is
unphysical for the tracer populations to have substantial rotation. In
the limit of mild or no rotation, the mass profile and anisotropy are
unaffected by the odd part of the distribution function. To this end,
we proceed under the assumption that the tracers are dominated by
random rather than systematic motion. In Table~\ref{tab:parent}, we estimate the fractional kinetic energy in rotation, $\left(V_{\rm
  rot}/\sigma\right)^2$, from the kinematic information given in the
literature for each sample of tracers (e.g. Table 7 in \citealt{coccato09}). In most cases,
this fraction is small ($< 20\%$) so it is safe to ignore rotation in
our analysis and assume spherical symmetry. However, in a few cases
(NGC 4564, NGC 4649 (GC) and NGC 5128 (PNe)), the rotation is quite
significant and our assumptions may not be valid. This is most
apparent for NGC 4564 where $V_{\rm rot}/\sigma \sim 1$. It is beyond
the scope of this paper to model the tracer populations with oblate or
triaxial spatial distributions, but we note that excluding these
systems from our sample has little difference on our main results.

\subsection{Line of sight velocity distribution}

In our sample of local galaxies, we do not possess full six-dimensional
phase space information for each tracer. However, we can easily
marginalise over the unknown components using the DF.  We marginalise
over the tangential velocity components ($v_l$ and $v_b$) and line of
sight distance ($D$) to obtain the line of sight velocity distribution
(LOSVD):
\begin{equation}
\label{eq:losvd}
F(l,b,v_{\rm los})=\int\int\int F(l,b,D,v_{l},v_{b},v_{\rm los})\,
\mathrm{d}v_l \, \mathrm{d} v_b \,\mathrm{d}D
\end{equation}
Here $l$, and $b$ are the Galactocentric longitude and latitude,
whilst $v_{\rm los}$ is the line of sight velocity.  We assume that
all the tracers are bound. Thus, the marginalisation over the
tangential velocity components and line of sight distance requires
$v_{\rm tot} < v_{\rm esc}$.

This method can be generalized to account for observational errors in
the line of sight velocities. We assume the errors are Gaussian and
marginalise over the line of sight velocity
\begin{equation}
\bar{F}(l,b,v_{\rm los})=\frac{1}{\sqrt{2 \pi \sigma_{\rm los,0}}}\int
F(l,b,v_{\rm los})
\mathrm{e}^{-\frac{\left(v-v_{los,0}\right)^2}{2\sigma_{\rm
      los,0}^2}}\mathrm{d}v_{\rm los}
\end{equation}
Here, $v_{\rm los,0}$ is the predicted line of sight velocity and
$\sigma_{\rm los,0}$ is the associated error.  There are four
parameters in our analysis: the potential normalisation ($\Phi_0$),
the potential power-law slope ($\gamma$), the velocity anisotropy
parameter ($\beta$) and the tracer density power-law slope
($\alpha$). We set the tracer density from the power-law
approximations made in the previous section. There remains three
parameters which we find using a maximum likelihood method.  The
likelihood function is constructed from the LOSVD
\begin{equation}
\label{eq:ml}
\mathrm{log} L(\beta, \Phi_0, \gamma)=\sum_{i=1}^N \mathrm{log}
\bar{F}(l_i,b_i,,v_{\mathrm{los}_i},\beta, \Phi_0, \gamma),
\end{equation}
Eqn~(\ref{eq:ml}) gives the three dimensional likelihood as a function of
$\beta$, $\gamma$ and $\Phi_0$. 
The total mass within a given radius can then be found from these parameters
\begin{equation}
M(< r)=\frac{\gamma \Phi_0}{\mathrm{G}}\left(\frac{r}{\mathrm{kpc}}\right)^{1-\gamma}.
\label{eq:mass}
\end{equation}
Here, $\Phi_0 = \Phi(1 \rm kpc)$. The analytic mass estimators of \cite{watkins10} also assume power-law
models for the potential and tracer density. Our analysis extends
beyond this formalism as, in addition to an estimate of the total
mass, we also constrain the slope of the potential and the velocity
anisotropy of the tracers. This is an important improvement as these
parameters are often poorly constrained both in theory and
observations. Furthermore, by constraining the slope of the potential
($\gamma$), we can report results at a common radius.

\begin{figure}
  \centering
  \begin{minipage}{\linewidth}
    \includegraphics[width=3.5in,height=2.5in]{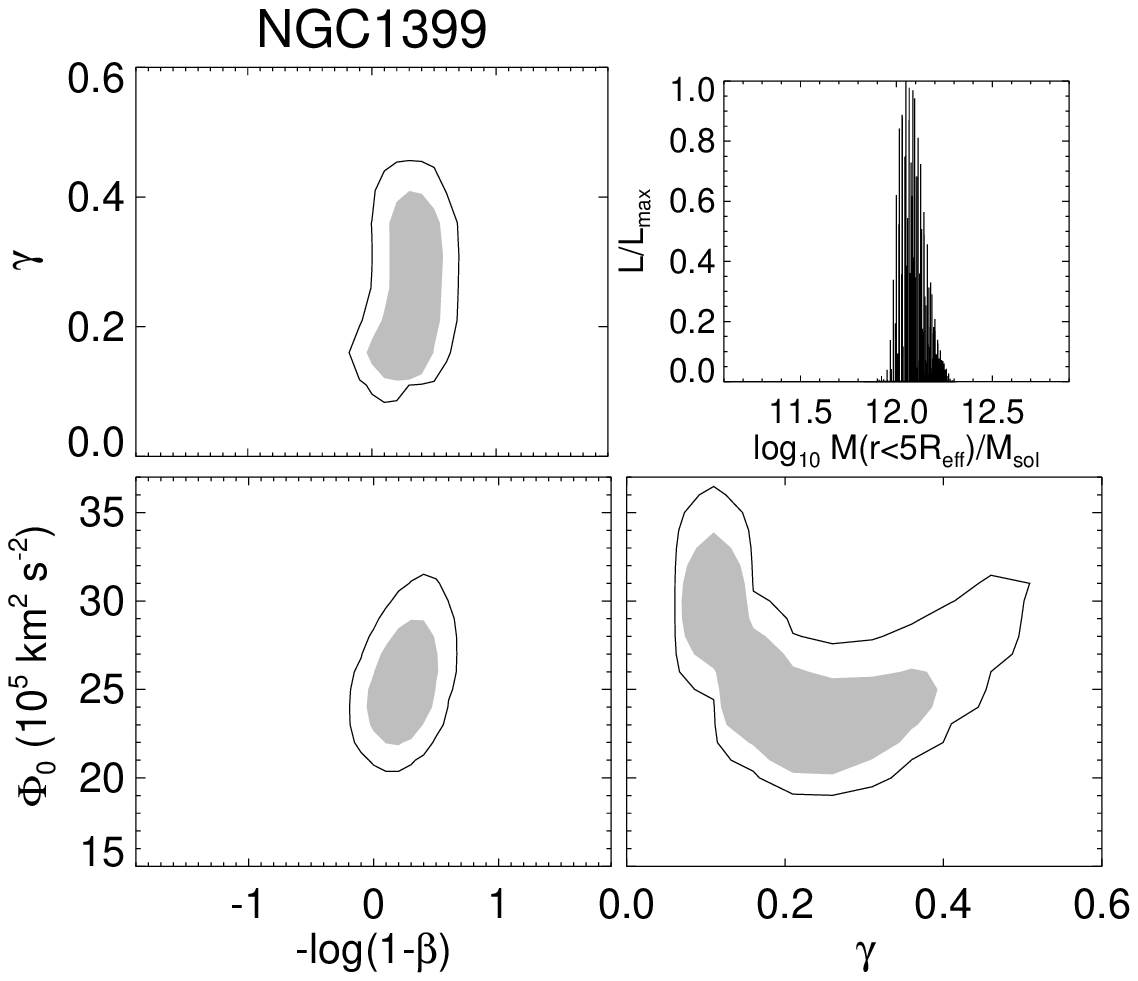}
  \end{minipage}
  \begin{minipage}{\linewidth}
    \includegraphics[width=3.5in,height=2.5in]{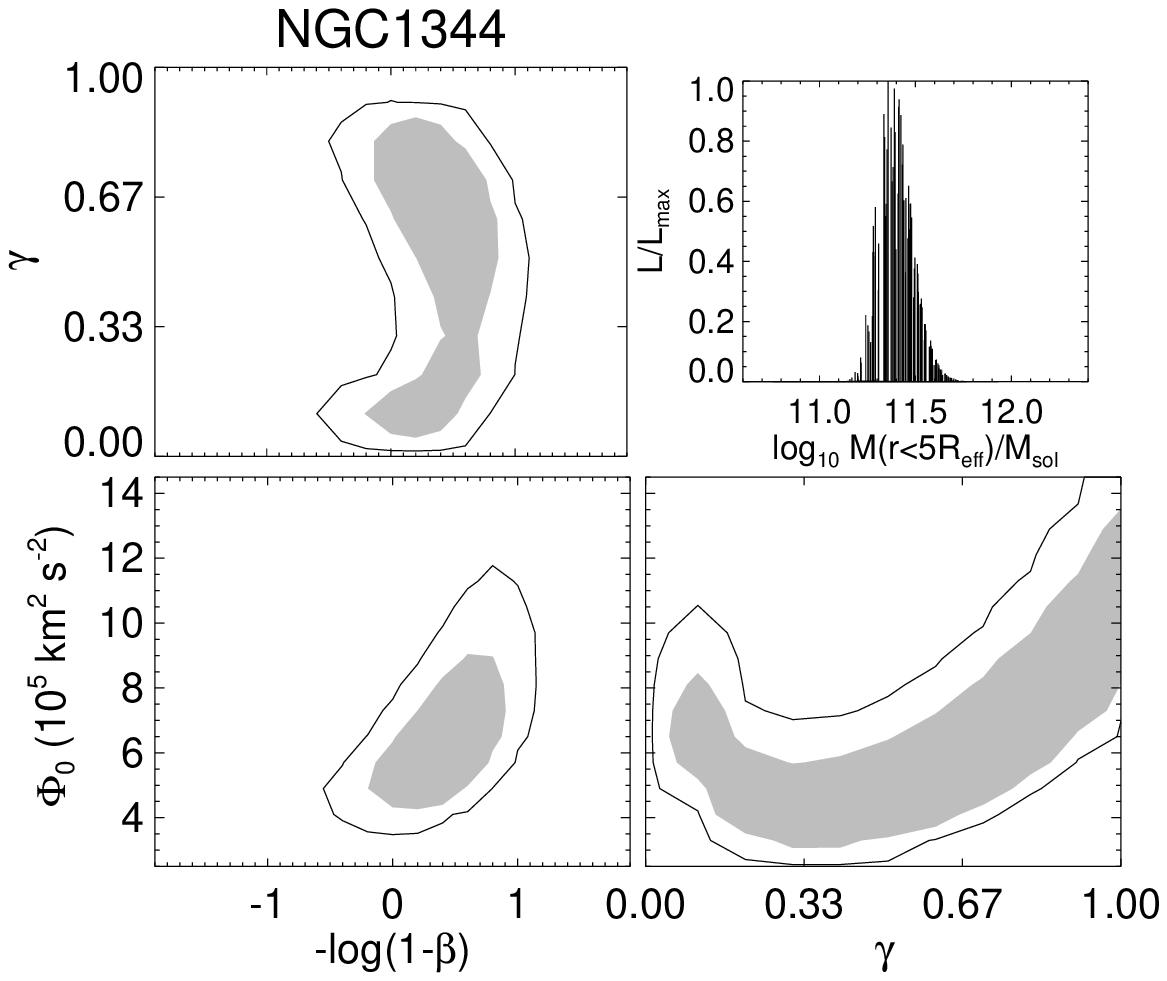}
  \end{minipage}
  \caption{The maximum likelihood contour levels for NGC 1399 and NGC 1344. The
      gray shaded regions show the $1\sigma$ (68\%) confidence region
      whilst the black lines encompass the $2\sigma$ (95\%) confidence
      region. The top right panel of each figure shows the likelihood
      as a function of total mass within $5R_{\rm eff}$.}
    \label{fig:contour}
\end{figure}

\begin{table*}
\begin{center}
\renewcommand{\tabcolsep}{0.1cm}
\renewcommand{\arraystretch}{1.3}
\begin{tabular}{| l c |c c c| c c c| c c c |}
\hline
Name & $R_{\rm max}/R_{\rm eff}$  & \multicolumn{3}{|c|}{$\beta$} & \multicolumn{3}{|c|}{$\Phi_0 [10^{5} \mathrm{km}^2 \mathrm{s}^{-2}]$} &  \multicolumn{3}{|c|}{$\gamma$} \\
(NGC)&  & $\bar{\beta}$ & $68\%$ & $95\%$ &  $\bar{\Phi}_0$ & $68\%$ & $95\%$ & $\bar{\gamma}$ & $68\%$ & $95\%$ \\
\hline
0821 & 8.1 & 0.2 & [0.1,0.5] & [-0.4,0.6] & 7.5 & [4.2,9.5] & [3.0,12.2] & 0.8 & [0.7,1.0] & [0.5,1.0] \\
1344 & 6.7 & 0.3 & [0.1,0.5] & [-0.3,0.6] & 7.0 & [3.5,8.4] & [3.0,11.9] & 0.6 & [0.4,1.0] & [0.1,1.0] \\
1399 & 22.2 & 0.2 & [0.1,0.4] & [-0.1,0.5] & 26.0 & [21.8,27.5] & [20.7,32.1] & 0.2 & [0.1,0.3] & [0.1,0.4] \\
1407 & 11.7 & -0.2 & [-0.5,0.2] & [-1.1,0.4] & 16.0 & [12.3,17.8] & [10.8,21.4] & 0.2 & [0.1,0.2] & [0.0,0.4] \\
3377 & 9.8 & 0.2 & [0.0,0.5] & [-0.4,0.6] & 2.6 & [1.6,3.2] & [1.2,4.5] & 0.3 & [0.1,0.3] & [0.0,0.8] \\
3379 & 9.2 & -0.0 & [-0.3,0.3] & [-1.0,0.4] & 4.1 & [2.7,4.9] & [2.5,6.1] & 0.7 & [0.5,1.0] & [0.3,1.0] \\
4374 & 5.7 & 0.3 & [0.2,0.4] & [-0.1,0.5] & 28.8 & [17.8,30.8] & [16.5,50.9] & 0.3 & [0.0,0.3] & [0.0,0.6] \\
4486 & 6.0 & -1.1 & [-1.9,-0.1] & [-4.6,0.3] & 42.5 & [29.2,48.1] & [26.0,72.4] & 0.4 & [0.2,0.7] & [0.1,0.8] \\
4494 & 7.4 & 0.1 & [-0.2,0.4] & [-0.5,0.5] & 3.5 & [1.9,3.9] & [1.6,5.7] & 0.7 & [0.6,1.0] & [0.3,1.0] \\
4564 & 10.5 & -1.1 & [-2.1,0.2] & [-6.4,0.5] & 1.6 & [0.9,1.7] & [0.8,2.6] & 0.7 & [0.7,1.0] & [0.1,1.0] \\
4636 & 8.6 & 0.2 & [0.0,0.5] & [-0.4,0.7] & 15.8 & [9.5,17.8] & [8.3,26.3] & 0.4 & [0.1,0.5] & [0.0,0.7] \\
4649 (GC) & 5.3 & -0.7 & [-1.5,0.2] & [-3.4,0.6] & 14.0 & [7.2,15.8] & [6.2,31.6] & 0.5 & [0.1,0.7] & [0.1,0.9] \\
4649 (PNe) & 4.1 & 0.1 & [-0.2,0.4] & [-0.6,0.5] & 18.0 & [6.9,21.7] & [6.5,33.5] & 0.6 & [0.6,1.0] & [0.0,1.0] \\
4697 & 5.2 & -0.5 & [-1.0,-0.1] & [-1.5,0.1] & 4.0 & [2.5,4.4] & [2.3,6.0] & 0.7 & [0.6,1.0] & [0.2,1.0] \\
5128 (GC) & 5.4 & 0.2 & [-0.1,0.4] & [-0.4,0.5] & 9.2 & [6.2,10.4] & [5.3,13.4] & 0.4 & [0.1,0.5] & [0.0,0.8] \\
5128 (PNe) & 15.6 & 0.5 & [0.4,0.6] & [0.3,0.7] & 10.7 & [6.1,12.6] & [5.0,16.8] & 0.7 & [0.6,0.9] & [0.4,1.0] \\
5846 & 3.9 & 0.2 & [0.0,0.5] & [-0.7,0.7] & 23.8 & [8.2,28.1] & [6.8,53.2] & 0.6 & [0.6,1.0] & [0.1,1.0] \\
\hline
\end{tabular}
\end{center}
\caption{Likelihood parameters.  We give the galaxy name and maximum
  projected radius (scaled by effective radius) in the first two
  columns. The remaining columns give the weighted mean of the velocity anisotropy, potential
  normalisation and potential power-law slope with their 68\% and 95\%
  confidence intervals.}
\label{tab:ml}
\end{table*}

\section{Results}

We summarise the maximum likelihood parameters in
Table~\ref{tab:ml}. In Fig.~\ref{fig:contour}, we show the likelihood
contours for NGC 1399 and NGC 1344 as examples. The grey shaded region shows the
$68\%$ confidence boundary and the solid line gives the $95\%$
confidence boundary. Note that each panel shows a 2D slice of the
likelihood values. There is a strong degeneracy between the potential
normalisation and power-law slope (note the `banana' shape in the
bottom right-hand panels). While we do not strongly constrain these individual
parameters, the mass profile is better defined as shown in the top
right hand panels of Fig.~\ref{fig:contour}. 

First, we compare our results to the analytic mass estimators described
in \cite{watkins10}. These mass estimators compute the mass within the
maximum 3D radius of the tracers, $r_{\rm max}$. For a spherical
distribution of stars with density distribution $\rho$, the average
  3D radius of a star with projected radius, R is given by:
\begin{equation}
\langle r \rangle =\frac{\int_{-\infty}^{\infty}r \, \rho(R,z) \,
  dz}{\int_{-\infty}^{\infty}\rho(R,z) \, dz}
\end{equation}
For a power-law distribution of tracers, this equation is analytic and
can be expressed as
\begin{equation}
\langle r \rangle
=\frac{\Gamma[\alpha/2]\Gamma[\alpha/2-1]}{\left(\Gamma[\alpha/2-1/2]\right)^2}
\, R
\end{equation}

\begin{figure}
  \centering
    \includegraphics[width=3.4in,height=2.267in]{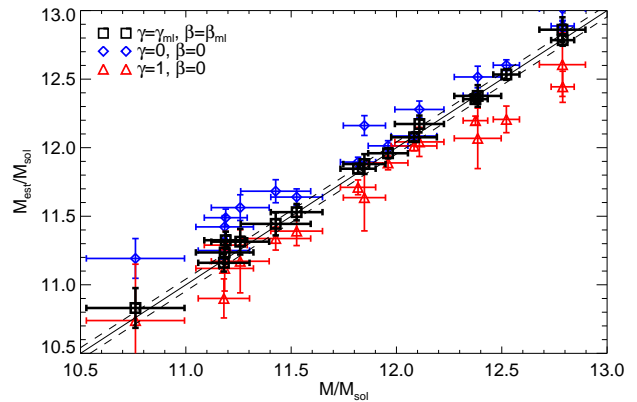}
    \caption[0]{Comparison with the mass estimator of
      \cite{watkins10}. We show the agreement between our
      masses within $r_{\rm max}$ ($M(r<r_{\rm max})$) and the
      estimated masses from the \cite{watkins10} formalism ($M_{\rm
        est}$). The black squares show the
      \cite{watkins10} mass estimates given our maximum likelihood
      $\beta$ and $\gamma$. The red triangles and blue diamonds show the
      mass estimators assuming velocity isotropy ($\beta=0$) with a
      Keplerian potential ($\gamma =1$) and an isothermal potential
      ($\gamma=0$) respectively. The solid line shows a one-to-one
      relation, and the dashed lines show 10\% discrepancies.}
    \label{fig:mest}
 \end{figure}

Here, $\alpha$ is the power-law index of the tracer density
profile. Using equations 26 and 27 from \cite{watkins10}, we estimate
the mass using the maximum likelihood $\gamma$ and $\beta$
values\footnote{Note that \cite{watkins10} label the power-law
  potential index and tracer density index as $\alpha$ and $\gamma$
  respectively. This is the opposite to the notation adopted in this
  work.}. We also evaluate the mass assuming an isothermal potential
($\gamma = 0$)\footnote{In the case $\gamma=0$, the potential
    is logarithmic (e.g. see equation 9 of \citealt{watkins10}). This
    is the limit of the power-law models as $\gamma \rightarrow 0$.}
or Keplerian potential ($\gamma = 1$) with isotropic orbits
($\beta=0$). These values are compared to our maximum likelihood mass
within $\langle r_{\rm max} \rangle$ (see eqn~\ref{eq:mass}) in
Fig. \ref{fig:mest}. 

We see very good agreement between
our estimated masses and those of \cite{watkins10} when the potential
slope and anisotropy are given. The agreement is within $\sim 10\%$
for all the systems. However, when an isothermal potential (blue) or
Keplerian potential (red) with isotropic tracers is assumed, there can
be quite large discrepancies. The mass is systematically overestimated
when an isothermal potential is assumed. In this case, the
discrepancies can be as large as $100\%$. Conversely, if we had
adopted a Keplerian potential the mass would be systematically
underestimated. The disagreement is worsened for systems where the
velocity anisotropy deviates from isotropy and/or the potential is not
close to isothermal or Keplerian. These findings illustrate the
effectiveness of our technique to estimate masses, as we make no
strong assumptions about the velocity anisotropy or potential
power-law slope.

\begin{figure}
  \centering
    \includegraphics[width=3.4in,height=2.267in]{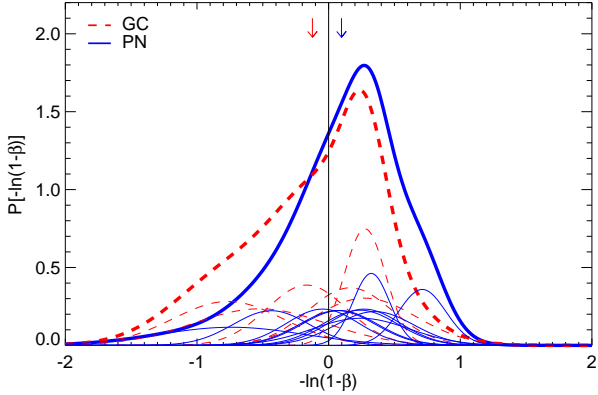}
    \caption{The velocity anisotropy distributions for the globular
      clusters (red dashed lines) and PNe (blue solid lines). The thick lines show
      the sum of the Gaussian kernels for each measurement. Individual
      contributions are shown with the thinner lines. In general, the
      PNe orbits are isotropic/mildly radially biased whilst the
      globular cluster orbits are isotropic/mildly tangential.}
    \label{fig:beta}
 \end{figure}

\begin{figure}
  \centering
    \includegraphics[width=3.4in,height=2.267in]{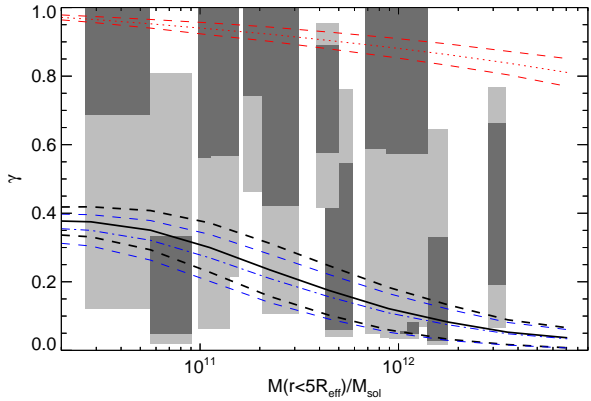}
    \caption[]{The slope of the overall power-law potential ($\Phi
      \propto r^{-\gamma}$) vs. the total mass enclosed within 5
      effective radii. The dark and light gray bands show the 68\% and
95\% confidence regions respectively. The solid black, dot-dashed blue
and dotted red lines
give the model predictions (within $10-100$kpc) for bulge+halo, halo
and bulge models respectively.}
    \label{fig:slope}
 \end{figure}

In Fig. \ref{fig:beta}, we show the velocity anisotropy distributions
of the globular cluster tracers (red lines) and PNe tracers (blue
lines). Each measurement is described by a Gaussian centred on the estimated value with a dispersion given by the predicted ($1 \sigma$) error. These Gaussian kernels are then summed to produce an overall distribution which is shown by the
  thick lines. Individual contributions are shown with the thinner
lines.  The PNe are generally more radially biased than the globular
clusters. This has been noted by previous authors who find that
GCs generally have mildly tangential/isotropic orbits
whilst PNe tend to have mildly radial/isotropic orbits
(e.g. \citealt{hwang08}; \citealt{romanowsky09}; \citealt{woodley10};
\citealt{napolitano11}). Note that the most tangentially biased PNe
system belongs to NGC 4564 which, as noted in the previous section,
has evidence for substantial rotation.

The slope of the overall power-law potential ($\Phi \propto
r^{-\gamma}$) is shown in Fig. \ref{fig:slope} as a function of total
mass within five effective radii\footnote{The total mass is computed
  within a 3D distance of $5R_{\rm eff}$. The scaling relation between
  3D deprojected half-light radius ($r_{1/2}$) and 2D projected
  half-light radius is $\sim 1.3$ (e.g. \citealt{ciotti}). Hence, a
  spherical distance of $5R_{\rm eff}$ corresponds to $\approx
  3.8r_{1/2}$.}. The black points give the weighted mean $\gamma$
values\footnote{The `weighted' mean is given by, $\bar{x}=\frac{\sum
    w_i x_i}{\sum w_i}$ where the weights are given by the likelihood values.} and the dark and light gray bands show the 68\% and
95\% confidence regions respectively. The solid and dashed lines give model
predictions for the power-law slope with their associated
scatter. These models are described in more detail in Section
\ref{sec:models}. The black lines show the slope-mass
relation for bulge+halo models where a Hernquist profile is assumed
for the stellar component and a Navarro-Frenk-White profile is assumed
for the dark matter component. The red and blue lines show the
relation with only a stellar or dark matter component
respectively. The slopes for these models are computed between
$10-100$ kpc.

At present, our $\gamma$ values are too poorly constrained to
distinguish between different model profiles. In the 95\% confidence
interval the slopes lie in between an isothermal ($\gamma=0$) and
Keplerian ($\gamma=1$) regime. However, there is tentative
  evidence for a trend with galaxy mass -- namely, the less massive
  galaxies have steeper potential profiles than the more massive
  galaxies. This trend is clear in the models, but is less so in the
  data. We tested the (anti)-correlation of $\gamma$ with mass in the
  data using a Spearman rank test. Given the probability distribution
  of $\gamma$ values (derived from the marginalised likelihood
  distributions), we draw a $\gamma$ value at random for each
  galaxy. The correlation with mass is then tested with a Spearman
  rank statistic. The exercise is repeated for $10^5$ trials. We find
  that only $20\%$ of the trials have a $95\%$ significant trend. This
  exercise suggests that it is premature to claim an
  (anti)-correlation between potential slope and mass from the current
  constraints. We note that \cite{barnabe11} also find a tentative
  indication that less massive galaxies have steepr potential slopes.

\cite{auger10b} and \cite{koopmans09} find almost isothermal profiles
for the SLACS elliptical galaxy sample within $1R_{\rm eff}$. The
SLACS sample is biased towards the more massive galaxies where
isothermal models are also predicted by the bulge+halo models.

\subsection{Dark matter fractions}
\label{sec:dm}

The maximum likelihood normalisation (i.e. total mass) and power-law
slope define the \textit{overall} potential. To separate the dark
matter and baryonic components, we assume the stars follow a Hernquist
profile (\citealt{hernquist90}). To convert luminosity into stellar
mass, we assume either a Chabrier/KTG (\citealt{chabrier03};
\citealt{kroupa93}) IMF with $(M/L_B)^* =3.5-5.5$ or a Salpeter
(\citealt{salpeter55}) IMF with $(M/L_{B})^* =6-10$
(e.g. \citealt{gerhard01}; \citealt{tortora09})\footnote{Note that
  stellar masses are larger by a factor of $\sim 1.8$ when a Salpeter
  IMF is adopted rather than a Chabrier/KTG IMF}. We assume
  the stellar mass-to-light ratios cover the given range with a flat
  prior, so all values in the range have equal weight. This range is
  taken into account in the estimated error of the stellar mass (see
  Table \ref{tab:nfw}). The fraction of dark matter within a certain
radius is given by
\begin{equation}
f_{\rm DM}(<r)=1-M^*(<r)/M_{\rm tot}(<r).
\end{equation}

Here, we have have assumed that the gas mass is negligible relative to
the stellar mass. As we are considering radial scales of $\sim
0.1r_{\rm vir}$ this is a good approximation
(e.g. \citealt{osullivan07}).

In recent years, the fraction of dark matter as a function of galaxy
mass has been studied extensively in the literature
(e.g. \citealt{padmanabhan04}; \citealt{cardone09};
\citealt{tortora09}; \citealt{cardone10}; \citealt{auger10b};
\citealt{grillo10}; \citealt{barnabe11}). Most of these studies have been limited to within
one effective radius. Here, we concentrate on the fraction of dark
matter within five effective radii - a region relatively unexplored in
local elliptical galaxies.

\subsubsection{Models}
\label{sec:models}

\begin{table}
\begin{center}
\renewcommand{\tabcolsep}{0.05cm}
\renewcommand{\arraystretch}{1.}
\begin{tabular}{| c c c c c c |} 
\hline
Name & $M(< 5R_{\rm eff})$ & $M/L_{B}$ & IMF & $M^*$  &$f_{\rm DM}$ \\ 
(NGC)&[$10^{11} M_{\odot}]$ & $(<5R_{\rm eff})$& & [$10^{11} M_{\odot}]$ &$(< 5R_{\rm eff})$\\

\hline
0821 & 2.3 $\pm$ 0.6 & 8 $\pm$ 2 & C & 1.2 $\pm$ 0.2 & 0.59 $\pm$ 0.11 \\
 &  & & S & 2.1 $\pm$ 0.4 & 0.27 $\pm$ 0.20 \\
\hline
1344 & 2.6 $\pm$ 0.5 & 14 $\pm$ 2 & C & 0.8 $\pm$ 0.2 & 0.74 $\pm$ 0.05 \\
 &  & & S & 1.5 $\pm$ 0.3 & 0.54 $\pm$ 0.10 \\
\hline
1399 & 12.5 $\pm$ 1.8 & 35 $\pm$ 5 & C & 1.6 $\pm$ 0.3 & 0.90 $\pm$ 0.02 \\
 &  & &S & 2.8 $\pm$ 0.5 & 0.82 $\pm$ 0.03 \\
\hline
1407 & 9.4 $\pm$ 1.3 &19 $\pm$ 2 & C & 2.1 $\pm$ 0.4 & 0.82 $\pm$ 0.03 \\
 &  & &S & 3.8 $\pm$ 0.7 & 0.67 $\pm$ 0.05 \\
\hline
3377 &0.7 $\pm$ 0.2 & 12 $\pm$ 2 & C & 0.3 $\pm$ 0.0 & 0.70 $\pm$ 0.07 \\
 &  & &S & 0.5 $\pm$ 0.1 & 0.46 $\pm$ 0.13 \\
\hline
3379 & 1.3 $\pm$ 0.2 &9 $\pm$ 1 & C & 0.6 $\pm$ 0.1 & 0.61 $\pm$ 0.07 \\
 &  & &S & 1.1 $\pm$ 0.2 & 0.31 $\pm$ 0.12 \\
\hline
4374 & 15.9 $\pm$ 1.9 &31 $\pm$ 3 & C & 2.2 $\pm$ 0.4 & 0.89 $\pm$ 0.01 \\
 &  & &S & 4.0 $\pm$ 0.7 & 0.80 $\pm$ 0.02 \\
\hline
4486 &31.7 $\pm$ 3.2 & 43 $\pm$ 4 & C & 3.3 $\pm$ 0.6 & 0.92 $\pm$ 0.01 \\
 &  & &S & 5.8 $\pm$ 1.0 & 0.85 $\pm$ 0.02 \\
\hline
4494 & 1.2 $\pm$ 0.2 &5 $\pm$ 1 & C & 1.0 $\pm$ 0.2 & 0.32 $\pm$ 0.12 \\
 &  & &S & 1.8 $\pm$ 0.3 & -0.21 $\pm$ 0.22 \\
\hline
4564 & 0.4 $\pm$ 0.1 &7 $\pm$ 2 & C & 0.2 $\pm$ 0.0 & 0.54 $\pm$ 0.17 \\
 &  & &S & 0.4 $\pm$ 0.1 & 0.18 $\pm$ 0.30 \\
\hline
4636 & 10.8 $\pm$ 1.9 &40 $\pm$ 6 & C & 1.2 $\pm$ 0.2 & 0.91 $\pm$ 0.02 \\
 &  & &S & 2.1 $\pm$ 0.4 & 0.84 $\pm$ 0.03 \\
\hline
4649 & 8.8 $\pm$ 1.3 &14 $\pm$ 2 & C & 2.8 $\pm$ 0.5 & 0.74 $\pm$ 0.04 \\
 & & &S & 4.9 $\pm$ 0.9 & 0.54 $\pm$ 0.07 \\
\hline
4697 & 1.4 $\pm$ 0.2 &7 $\pm$ 1 & C & 0.8 $\pm$ 0.2 & 0.51 $\pm$ 0.08 \\
 &  & &S & 1.5 $\pm$ 0.3 & 0.12 $\pm$ 0.13 \\
\hline
5128 & 4.7 $\pm$ 0.5 &17 $\pm$ 1 & C & 1.2 $\pm$ 0.2 & 0.79 $\pm$ 0.02 \\
 & & &S & 2.2 $\pm$ 0.4 & 0.63 $\pm$ 0.04 \\
\hline
5846 & 11.3 $\pm$ 2.7 &28 $\pm$ 6 & C & 1.8 $\pm$ 0.3 & 0.87 $\pm$ 0.03 \\
 &  & &S & 3.2 $\pm$ 0.6 & 0.77 $\pm$ 0.05 \\
\hline
 
 \end{tabular}
\end{center}
\caption{Dark matter fraction parameters. We give the galaxy name,
  total mass-to-light ratio within $5R_{\rm eff}$, adopted IMF
  (Chabrier/KTG (C) or Salpeter (S)), stellar mass and fraction of dark
  matter within $5R_{\rm eff}$.}
\label{tab:nfw}
\end{table}

Before proceeding, we construct some simple models to show the expected
relation between dark matter fraction (within a certain radius) and
galaxy mass. We apply the following steps:

\begin{itemize}

\item Abundance matching is used to relate the halo mass ($M_{200}$)
  to the stellar mass. We use the prescription given by
  \cite{behroozi10}. The stellar mass-halo mass relation is given by
  their Equation (21) with parameters listed in Table (2). These authors adopt a $\Lambda$CDM cosmology
  using the WMAP5 parameters. The scatter in stellar mass for a given
  halo mass is $\sim 0.15$ dex. This abundance matching is applicable
  for a Chabrier/KTG IMF. To convert to a Salpeter IMF, the stellar
  masses are increased by a factor of 0.25 dex.

\item An NFW profile is adopted to model the density profile of the
  halo. This profile is fully described by two parameters: the halo
  mass and concentration. We relate the concentration to the halo mass
  using the relation given in \cite{maccio08} for WMAP5. At this stage,
  we have a fully defined dark matter profile for a particular stellar
  mass. The scatter in this relation for a given halo mass is $\sim
  0.1$ dex.

\item Using the quadratic size-mass relation given in \cite{hyde09} based on
  SDSS data, we calculate the effective radius of a galaxy from the
  stellar mass. The scatter for a given stellar mass is $\sim 0.2$
  dex. The form of this size-mass relation is shown in
    the top panel of Fig,~\ref{fig:size_mass} against the observed values in our sample.

\item Finally, we adopt a Hernquist profile for the stellar profile
  which is defined from the total stellar mass and effective radius.

\end{itemize}

\begin{figure*}
  \centering
    \includegraphics[width=6.75in,height=2.25in]{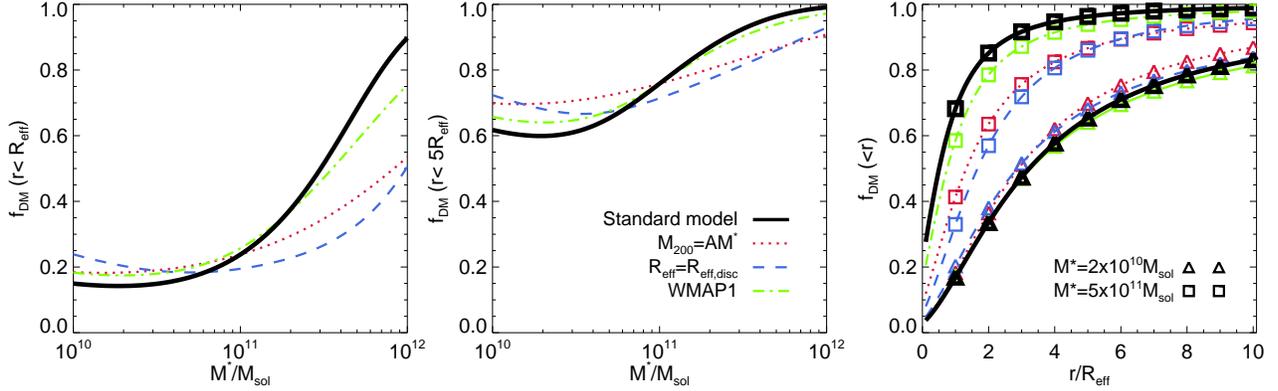}
    \caption[0]{The dark matter fraction within one (left panel) and
      five (middle panel) effective radii as a function of stellar
      mass. The solid black line is our adopted model. The dotted red line shows
      the relation when a linear stellar mass-halo mass relation is
      used $M_{200}=AM^*$. The dashed blue line shows a model with a
      size-mass relationship applicable to disc galaxies. The
      dot-dashed green
      line shows the model prediction when a WMAP1 cosmology is
      adopted. In the right-hand panel we show the dark matter
      fraction as a function of radius (scaled by $R_{\rm eff}$) for
      the different models. The triangle and square symbols show the
      profiles for lower ($M^* \sim 2 \times 10^{10} M_{\odot}$) and
      higher ($M^* \sim 5 \times 10^{11} M_\odot$) mass galaxies
      respectively.}
    \label{fig:dm_comp}
 \end{figure*}

\begin{figure}
  \includegraphics[width=3.4in,height=3.4in]{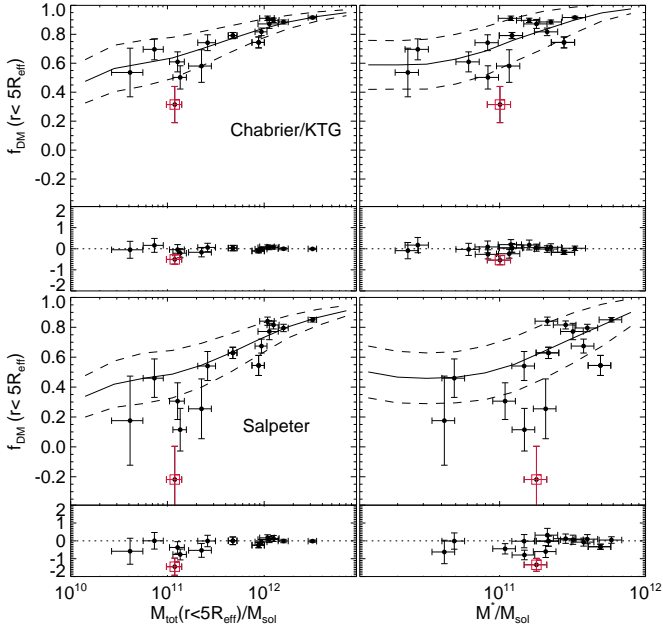}
  \caption{ The dark matter fraction ratio within 5 effective radii
    vs. total mass within 5 effective radii (left) and stellar mass
    (right). A Chabrier/KTG ($(M/L)^* \sim 4.5$) and Salpeter
    ($(M/L)^* \sim 8$) IMF is assumed in the top and bottom panels
    respectively. The solid line indicates the predicted relation from
    simulations. The predicted scatter is given by the dashed
    lines. The inset panels show the residuals of the data
    vs. models. The red square highlights NGC 4494 which is
    inconsistent with stellar mass-to-light ratios larger than
    $(M/L)^* \sim 5$.}
  \label{fig:dm_frac}
\end{figure}

We now have a dark matter and stellar profile defined for any
particular stellar mass. In addition, we take into account the scatter
introduced in each step of the analysis. These are propagated forward
using Monte Carlo techniques. We emphasize that there are no free
parameters in these models that need to be fitted to the data.

The models predict an increase in dark matter fraction within 5
$R_{\rm eff}$ with increasing total/stellar mass (see black line in
Fig. \ref{fig:dm_comp}). This general trend is independent of the IMF
used. The fraction of dark matter within a scaled number of effective
radii is driven by two factors: the star formation efficiency and the
concentration of these stars within the dark matter halo
(cf. \citealt{zaritsky08}).

The star formation efficiency is implemented in the models from the
stellar mass-halo mass relation. The abundance matching, by
construction, reproduces the observable stellar mass function. It is
well known that there is a universal U-shaped trend of star formation
efficiency (e.g. \citealt{benson00}; \citealt{marinoni02};
\citealt{napolitano05}; \citealt{vandenbosch07};
\citealt{conroy09}). The peak efficiency occurs at $M^* \sim 10^{11}$
where increasingly massive galaxies have a lower efficiency due to the
large cooling time of their hot gas (e.g. \citealt{white78}) and the
least massive systems are unable to retain their primordial gas
content for long enough to form stars. Hence, the the lowest mass and
highest mass galaxies are the most dark matter dominated.

The second factor driving the trend of dark matter fractions is the
size-mass relationship of early type galaxies. With increasing total
(or stellar) mass the effective radius increases (see
Fig. \ref{fig:size_mass}). In addition, the size increases
\textit{more} steeply with increasing mass. This `baryon un-packing' is an
important factor when we are considering the dark matter content
within a scaled number of effective radii. Regardless of the dark
matter halo properties (or star formation efficiency), a higher
concentration of baryons (or smaller effective radius) leads to
smaller dark matter fractions. This simple scaling behaviour has been
noted by previous authors (e.g. \citealt{padmanabhan04};
\citealt{tortora09}; \citealt{napolitano10}; \citealt{dutton10}). Note
that disc galaxies have a much shallower size-mass relation. Thus, the
dark matter fractions are approximately constant with mass for these
galaxies (e.g. \citealt{dutton10}). 

In Fig.~\ref{fig:dm_comp}, we illustrate the dependence of our model on
the adopted stellar mass-halo mass relation, size-mass relation and
cosmology. The solid black lines indicate our adopted model. We consider
three variations to our adopted model (shown by the coloured dotted,
dashed and dot-dashed lines):\\

\begin{itemize}

\item \textit{Stellar mass-halo mass relation}: We adopt a linear
  relation of the form $M_{200}=AM^*$. The normalisation constant is
  set using the mean halo mass corresponding to a stellar mass of $M^*
  \sim 10^{11}$ from our original abundance matching relation. Models
  adopting this relation are shown with the red lines.

\item \textit{Size-mass relation}: We consider a size-mass relation
  applicable to disc galaxies. This is derived using the
  \cite{pizagno07} sample of disc galaxies. The stellar masses are
  derived from the colours according to the prescription of
  \cite{bell03}. A -0.15dex offset was applied to provide stellar
  masses consistent with a Chabrier/KTG IMF. We then fit a linear
  relation between effective radius and stellar mass to approximate a
  size-mass relationship. Models assuming this size-mass relation are
  shown by the blue lines.

\item \textit{Cosmology}: We adopt a WMAP1 cosmology instead of the
  more recent WMAP5 parameters.  In this case the stellar mass-halo
  mass relation is taken from \cite{guo10} (who adopt a WMAP1
  cosmology) and the WMAP1 mass concentration relation is used from
  \cite{maccio08}. This is shown by the green lines.

\end{itemize}

The purpose of these variations is to emphasize the key ingredients in
our adopted model driving the apparent trend. Thus, these variations
are for illustrative purposes only and we are not proposing that they
are viable alternatives to our adopted model. For example, adopting a
size-mass relation applicable to disc galaxies is a particularly poor
assumption when applied to elliptical galaxies.

The left and middle panels show the dark matter fraction within one
and five effective radii as a function of stellar mass. We have not
included the scatter, so this figure only illustrates the mean
dependence of the models. Note that a Chabrier/KTG IMF is assumed but
the same trends are seen with a Salpeter IMF.  Adopting a different
cosmology only slightly affects the dark matter fraction-stellar mass
relation. As alluded to earlier, the relationship is driven by star
formation efficiency and baryon extent. By adopting a shallower
size-mass relation (applicable to disc galaxies) the dark matter
fractions are constant over a large range of stellar mass. A similar
effect is seen by allowing for a linear stellar mass-halo mass
relation. The effects are the most severe within one effective radii
as the baryons are more dominant at smaller radii.

In the right-hand panel, we show the dark matter fraction as a
function of radius (scaled by $R_{\rm eff}$). The profiles of the
higher mass galaxies ($M^* \sim 5 \times 10^{11} M_\odot$) are more
model dependent than those of lower mass galaxies ($M^* \sim 2 \times
10^{10} M_\odot$).

\subsubsection{Comparison with data}

In Fig.~\ref{fig:dm_frac}, we show the dark matter fractions within
five effective radii as a function of total mass within this radius
(left) and as a function of total stellar mass (right). The top panels
show the relation for a Chabrier/KTG IMF and bottom panels are for a
Salpeter IMF. The symbols with error bars are the data points derived
in this work and the solid and dashed lines give the predicted
relation from the models with the associated ($1\sigma$) scatter. The
residuals of the data vs. model (i.e. (data-model)/model) are shown in
the inset panels.

In general, the model is in good agreement with the data when a
Chabrier/KTG IMF is adopted. Our dark matter fractions are in good
agreement with previous studies which find $f_{\rm DM}(<5R_{\rm eff})
\sim 0.4-0.5$ for ordinary ellipticals (e.g. NGC 4494, NGC 3379, NGC
4697 see Figure 12 in \citealt{napolitano11}) and $f_{\rm DM}(<5R_{\rm
  eff}) \sim 0.8-0.9$ for group or cluster central ellipticals
(e.g. NGC 1399, NGC 1407, NGC 4486, NGC 4648, NGC 5846 see Figure 8 in
\citealt{das10}).

Note that our total mass estimates (given within $5R_{\rm eff}$ in
Table \ref{tab:nfw}), and hence dark matter fractions, usually have
smaller uncertainties than those found in more general dynamical
modelling (e.g. \citealt{delorenzi07}; \citealt{delorenzi09}). This is
because such numerical modelling procedures allow for more general
solutions (e.g. non-constant anisotropy, triaxiality) and thus cover a
larger parameter space. Nonetheless, our simplistic approach is much
more easily applied to a large sample of galaxies and hence this work
is complementary to more general numerical methods.

The red square highlighted is NGC 4494, which has a curiously low dark
matter fraction. \cite{napolitano09} studied this system in detail and
found an abnormally low dark matter concentration and also derive a
similar dark matter fraction within 5$R_{\rm eff}$. The remaining low
mass galaxies have dark matter fractions consistent with the model
predictions (at least for a Chabrier/KTG IMF). While the case of NGC
4494 is in contradiction to the predictions from simulations, the
general trend suggests that the dark matter properties of lower mass
galaxies are not very different from those of higher masses. However,
a larger sample of tracers, probing both low and high masses, is
needed to investigate this further.

Adopting a Salpeter IMF causes some of the low mass systems to have
lower than predicted dark matter fractions. On the other hand, a
Chabrier/KTG IMF is able to describe the trend over a wide range of
masses. A secondary factor that may influence the dark matter
fraction-mass relation is variations in stellar population properties,
such as a non-universal IMF (e.g \citealt{auger10a};
\citealt{treu10}). We find no evidence for a variation in IMF with
mass from Fig. \ref{fig:dm_frac}. The suggested IMF variation, where
the stellar-mass to light ratio increases with mass, would require the
fraction of dark matter to increase more steeply when a universal IMF
is adopted. However, we are probing out to five effective radii, well
beyond the regions where the baryonic material is dominant so are less
sensitive to IMF variations than studies probing within one effective
radii. We compare our models with such studies in the following
section.

\begin{figure}
  \includegraphics[width=3.4in,height=3.4in]{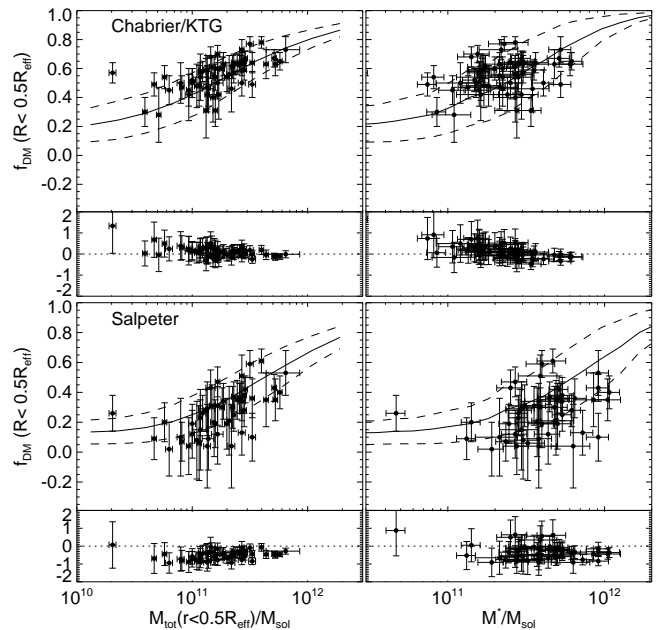}
  \caption{The projected dark matter fraction ratio within 0.5
    effective radii vs. total mass within 0.5 effective radii (left)
    and stellar mass (right). The data points are for the SLACS sample
    of elliptical galaxies published in \cite{auger10b}. The inset
    panels show the residuals of the data vs. model.}
   \label{fig:auger}
\end{figure}

\subsubsection{Comparison with SLACS data}

In Fig.~\ref{fig:auger}, we show the \textit{projected} dark matter
fractions as a function of galaxy mass and stellar mass. The data
points derive from the SLACS sample of \cite{auger10b}. This work
focuses on relatively high mass galaxies within one effective
radius. The simple models that we employ here are able to explain the
overall trend of increasing dark matter fraction with galaxy
mass. Similar to the case at larger radii, adopting a Salpeter IMF
leads to lower dark matter fractions than the model predictions. In
fact, several systems are consistent with a negative dark matter
fraction, which suggests that adopting a Salpeter IMF for such systems
is unphysical. Once again a \textit{universal} Chabrier/KTG IMF is
able to explain the dark matter fraction-mass relation reasonably
well. Variations in IMF with mass can in principle be probed by
deviations of the data from the model predictions. Qualitatively, we
see no evidence for a difference in slope between the data and the
model. However, a more sophisticated model is required to investigate
these variations in detail. In summary, we find that the variation in
dark matter fraction with galaxy mass is consistent with a simple,
universal IMF halo model.

Recently, \cite{grillo10} finds a constant, if not \textit{decreasing},
projected dark matter fraction with galaxy mass. This is inconsistent
with both our models and our measurements as well as with the
measurements by \cite{auger10b}, independently of the adopted
IMF. \cite{grillo10} suggests that these contrasting results may be due
to other studies being more model dependent
(e.g. \citealt{padmanabhan04}; \citealt{tortora09} who give three
dimensional dark matter fractions) and/or the more massive early type
galaxy bias of their sample. However, this is difficult to reconcile
with the results of \cite{auger10b} who also give projected dark
matter fractions (i.e. with minimal modelling assumptions) and have a
similar mass bias in their sample.

\section{Conclusions}

We studied the mass profiles of local elliptical galaxies using
planetary nebulae (PNe) and globular clusters (GCs) as distant tracers
(with $R > 2R_{\rm eff}$). A sample of 15 galaxies was compiled from
the literature. A distribution function-maximum likelihood method was
used to study the dynamics of the tracers under the assumptions of
spherical symmetry and power-law models for the (overall) potential
and tracer density. We summarise our conclusions as follows:

\bigskip
\noindent
(1) We compared our distribution function-maximum likelihood method to
the analytic mass estimators of \cite{watkins10}. There is good
agreement when we adopt our maximum likelihood velocity anisotropy and
potential power-law indices into the \cite{watkins10}
formalism. However, assuming isotropy and an isothermal/Keplerian
potential leads to a systematic overestimate/underestimate of the mass
(by up to $\sim 100\%$). Our method makes no strong assumptions about
the values of the velocity anisotropy\footnote{However, we do assume that $\beta$ is
  constant with radius} and potential power-law slope. This is an
important improvement, as these parameters are difficult to constrain
by other means.

\bigskip
\noindent
(2) The PNe are generally more centrally concentrated (following
closely the stellar luminosity, see \citealt{coccato09}) and are on
more radial orbits than the GCs. This may be related to the parent
galaxy properties as the PNe tend to trace the less massive systems,
but could also reflect differences between the GC and PNe populations
(e.g. GCs are generally older).

\bigskip
\noindent
(3)  The slope of the overall (power-law) potential is not strongly
constrained but lies in between the isothermal ($\gamma=0$) and Keplerian
($\gamma =1$) regimes. In the models the less massive galaxies have steeper
potential profiles than those of the more massive galaxies. There is
tentative evidence that this trend is also present in the
data. However, a Monte Carlo test using the Spearman rank statistic
shows no evidence for a clear trend. Recently,
\cite{koopmans09} and \cite{auger10b} found roughly isothermal density
profiles for their sample of (massive) elliptical galaxies within an
effective radius. This is in good agreement with bulge+halo models for
the more massive elliptical galaxies.

\bigskip
\noindent
(4) We constructed simple halo models to predict the expected relation
between dark matter fraction and halo mass. The combination of star
formation efficiency (stellar mass-halo mass relation) and baryon
concentration (size-mass relation) is able to describe the observed
trend of increasing dark matter fraction within a scaled number of
effective radii. We find no evidence for an additional factor, such as
a non-universal IMF, affecting this relation. The dark matter
fractions are consistent with the models when a \textit{universal}
Chabrier/KTG IMF is assumed for elliptical galaxies. In particular, a
Salpeter IMF is inconsistent with some of the lower mass
galaxies. This is in good agreement with \cite{cappellari06}
  who reached the same conclusion using a
completely different approach.

\bigskip
\noindent
Finally, let us remark that we have neglected the influence of
baryonic processes on the distribution of dark matter. Our adopted NFW
profile is applicable for a `pristine' dark matter distribution,
un-modified by the process of galaxy formation. For example,
collapsing gas can exert a gravitational drag on the dark matter
leading to a more concentrated dark matter halo
(i.e. \citealt{blumenthal86}; \citealt{gnedin04}). On the other hand,
rapid supernova driven feedback processes can drive the dark matter
particles away from the centre of the galaxy
(e.g. \citealt{pontzen11}) leading to halo expansion. Halo contraction
or expansion can increase or decrease the fraction of dark matter
within a given radius. The importance of these processes is poorly
understood and we make not attempt to model them in this
work. However, we note that this is another secondary effect (in
addition to IMF variations) that should be explored with more
sophisticated modelling.

\section*{Acknowledgements}
AJD thanks the Science and Technology Facilities Council (STFC) for
the award of a studentship, whilst VB acknowledges financial support
from the Royal Society. We would like to acknowledge Matt Auger,
Gary Mamon and Michele Cappellari, as well as an anonymous referee,
for valuable comments.

\bibliography{mybib}

\end{document}